\documentstyle[multicol,aps,bezier,epsfig,amsmath,amssymb]{revtex}
\newcommand{\upd}{{\mathrm d}}
\renewcommand{\epsilon}{\varepsilon}

\begin{document}

\draft

\twocolumn[\hsize\textwidth\columnwidth\hsize\csname@twocolumnfalse\endcsname

\title{ \bf Phase separation in an homogeneous shear flow: \\
Morphology, growth laws and dynamic scaling}

\author{Ludovic Berthier}  

\address{
Laboratoire de Physique, \'Ecole Normale Sup\'erieure,
\\
46 All\'ee d'Italie,  F-69007 Lyon France and
\\
D\'epartement de Physique des Mat\'eriaux, U. C. B. Lyon 1,
\\
43 Bd du 11 Novembre 1918, 69622 Villeurbanne Cedex, France 
}  

\date{\today}

\maketitle

\begin{abstract}
We investigate numerically the influence of an homogeneous shear flow on the
spinodal decomposition of a binary mixture by solving 
the Cahn-Hilliard equation
in a two-dimensional geometry.
Several aspects of this much studied problem are clarified.
Our numerical data show unambiguously that, in the shear flow, the domains
have on average an elliptic shape.
The time evolution of the three parameters describing this ellipse
are obtained for a wide range of shear rates.
For the lowest shear rates investigated, we find the growth laws
for the two principal axis  
$R_\perp (t) \sim constant$, $R_\parallel(t) \sim t$, while the 
mean orientation of the domains 
with respect to the flow is inversely proportional to the strain.
This implies that when hydrodynamics is neglected a shear flow
does not stop the domain growth process.
We investigate also the possibility of dynamic scaling,
and show that only a non trivial form of scaling holds, 
as predicted by a recent analytical approach to the case of a 
non-conserved order parameter.
We show that a simple physical argument may account
for these results.
\end{abstract}  

\pacs{PACS numbers: 05.70.Ln, 47.55.Kf, 64.75.+g  \hspace*{4.3cm}
LPENSL-TH-14/2000}




\twocolumn\vskip.5pc]\narrowtext

\section{Introduction}

The study of phase ordering kinetics
has a long  history~\cite{review_hohenberg,review_gunton,review_alan}.
The canonical example is the coarsening process following 
the  quench of a binary mixture A-B below
its spinodal line. 
The properties of the resulting domain growth are rather well 
understood~\cite{review_hohenberg,review_gunton,review_alan}.
In the case where A and B are in equal concentration,
an isotropic bicontinuous structure emerges, which is 
characterized by a typical
length scale $L(t)$ growing as a power law of time $t$.
Moreover, as $t$ increases, the structure evolves 
in a self-similar manner in the sense that its
statistical properties are the same when space is rescaled by $L(t)$.
Any function $C(\boldsymbol{r},t)$ that depends on space and time
is then a function 
the reduced variable 
$|\boldsymbol{r}|/L(t)$ only:
$
C(\boldsymbol{r},t) \equiv {\cal C} \left(
|\boldsymbol{r}|/L(t) \right)$.
This property is termed `dynamic 
scaling'~\cite{review_hohenberg,review_gunton,review_alan}.
 
The study of the spinodal decomposition in an homogeneous shear
flow is of fundamental and practical interest~\cite{review_onuki}, but
despite an enormous amount of 
experimental~\cite{beysens0,beysens1,string,hobbie,qiu2,lauger,beysens2}, 
numerical~\cite{cogola,cates,shou,julia,zhang,yamamoto,cogola2,rothman,chan,qiu,padilla,ohta} and
analytical works\cite{review_onuki,cogola,rabr,brca,standrews}, 
the problem is still not fully settled~\cite{standrews}.
Early experiments~\cite{beysens0,beysens1,beysens2} and 
simulations~\cite{rothman,chan,ohta} have shown that 
isotropy is lost, the morphology 
of the bicontinuous structure being elongated along the flow direction.
Hence, a single length scale can not describe 
the full structure. 
There are thus several questions which naturally arise.
(1) How to characterize quantitatively the growing structure?
(2) What is the time dependence of the different length scales, 
as compared to the unsheared case?
(3) Does the growth stop after some time? 
(4) Does a suitable generalization of the dynamic scaling 
property hold?

It is interesting to remark that in spite of the large number of works cited
above, a definite answer to all these questions is still lacking.
Several reasons make this problem non-trivial.
First, the role of hydrodynamics is far from being understood.
Simple physical arguments~\cite{review_onuki} show that it may become
preponderant at large times, predicting 
a saturation of the domains at a typical size.
This assumption is resolved neither at the experimental level
(see, e.g., the opposite conclusions of Ref.~\cite{string} and 
Refs.~\cite{beysens2}) nor at the numerical level, where 
powerful algorithms are needed to correctly deal 
with hydrodynamics.
At present, this limits the numerical analysis to too small 
sizes to make any definite answers to questions (1)-(4) above, although
much progress has been made recently~\cite{cates,shou,julia,zhang}. 

Second, at the theoretical level, no analytical solution of
a reasonable model of spinodal decomposition (even neglecting
hydrodynamics) is available. 
One has then to make some predictions from the solution
of solvable, but less realistic models, like the $O(n)$ model
in the large-$n$ limit~\cite{rabr}, or from the approximate solution
in the case of a non-conserved order parameter~\cite{brca}.
A scaling argument based on the hypothesis that a  
generalization of dynamic scaling holds
has also been developed in Ref.~\cite{cogola}.

Third, there are, to our knowledge, no numerical simulations
(neglecting hydrodynamics) validating these analytical predictions.
Moreover, the scaling hypothesis on which is based the analytical
argument of Ref.~\cite{cogola} has only been tested in 
Ref.~\cite{qiu2}, with negative results. 
Thus, the validity of the predicted growth laws 
may also be questioned.
More crucially, there is up to now no consensus concerning
the numerically measured growth laws: we discuss this point in 
a more detailed way in the last Section.

In this work, we numerically study the spinodal 
decomposition process in a shear flow
by solving the Cahn-Hilliard equation in two dimensions. 
All hydrodynamic effects are neglected. 
Although this involves a drastic reduction of the experimental situation, 
it is necessary, in our opinion, 
to have a good understanding of this `simple' case, before studying 
more realistic problems.
Our algorithm is different of but our study is 
technically comparable to the most recent one~\cite{cogola}.
We shall however explore a wider range of shear
rates, and this will lead us to a different interpretation
of the numerical data.

The paper is organized as follows. In the next Section, we define
the model and describe the numerical procedure to solve it.
Section III briefly recalls the results obtained when
the shear flow is absent.
Section IV describes the morphology of the domains under shear and its
time evolution.
Section V focuses on the problem of dynamic scaling.
In the last Section, we compare our results with the relevant existing
data in the literature and give a simple physical argument 
to explain them.

\section{Model and details of the simulation}

In this work we focus on the standard model for
the spinodal decomposition of binary mixtures and solve 
numerically the Cahn-Hilliard 
equation~\cite{review_hohenberg,review_gunton,review_alan,review_onuki} 
\begin{equation}
 \frac{\partial \phi(\boldsymbol{r},t)}{\partial t} + \boldsymbol{v} \cdot 
\boldsymbol{\nabla} \phi(\boldsymbol{r},t)  =  \Gamma \nabla^2 \left( \frac
 {\delta F[\phi]}{\delta \phi 
(\boldsymbol{r},t)} \right) + \eta(\boldsymbol{r},t).
\label{cahn}
\end{equation}   
In this expression, the order parameter $\phi(\boldsymbol{r},t)$
is a scalar quantity which can be linked to the concentration $c_A$
($c_B$) of the component A (B) of the mixture by the relation
$\phi \equiv 1 - 2 c_A = 2c_B -1$.
Eq.~(\ref{cahn}) has the form of a continuity equation, which
implies that the order parameter is a conserved quantity.
The free energy $F[\phi]$ is of the Ginzburg-Landau type
\begin{equation}
F[\phi] = \int\nolimits \upd^d \boldsymbol{r} \left[ \frac{\xi^2}{2} 
{|\boldsymbol{\nabla}\phi|}^2 +
\frac{1}{4}\phi^{4} - \frac{1}{2} \phi^2  \right] ,
\label{landau}
\end{equation}
where the equilibrium correlation length $\xi$ has been introduced.  
The noise term $\eta$ is a random Gaussian variable, characterized 
the two moments $\langle\eta(\boldsymbol{r},t)\rangle = 0$
and $\langle\eta(\boldsymbol{r},t) \eta(\boldsymbol{r'},t')\rangle =
 -2T\delta(t-t') \nabla^2 \delta(\boldsymbol{r}-\boldsymbol{r'})$;
$T$ is the temperature of the thermal bath. 
All the simulations will be carried out at $T=0$, since temperature
is essentially irrelevant in this process~\cite{review_alan}.
It is also known to delay the onset of 
the asymptotic regime~\cite{rogers}.
The second term in the left hand side of Eq.~(\ref{cahn})
results from the advection
of the order parameter by the velocity field.
The case of an homogeneously sheared system will be 
investigated.
The flow is taken to be in the $x$-direction and the velocity
field is then
$\boldsymbol{v} = {\gamma} y \,\, \boldsymbol{e_x}$,
which defines the shear rate $\gamma$.
We shall concentrate on the case of a constant shear rate.

We want then to solve numerically in two spatial dimensions
the following equation:
\begin{equation}
\frac{\partial \phi}{\partial t} = - 
{\gamma} y \frac{\partial 
\phi}{\partial x} - \Gamma \nabla^2 \left( \xi^2 \nabla^2 \phi - \phi^3+\phi 
\right) , 
\label{cahn2}
\end{equation}
where both space and time dependences have been removed for clarity.
This is done by combining the numerical methods of Refs.~\cite{onuki,BBK1}.
A new frame ($x',y'$) is first defined by~\cite{onuki}
$x' \equiv x - S(t) y$, $y' \equiv y$, where 
$ S(t) \equiv \int_{0}^{t} \upd t' {\gamma}$ is the strain.
In the case of a constant shear rate, it is simply given by
$S(t) = \gamma t$.
Defining further
$\phi(\boldsymbol{r},t) \equiv \hat{\phi} 
(\boldsymbol{r'},t)$, Eq.~(\ref{cahn2}) becomes~\cite{onuki}
\begin{equation}
\frac{\partial \hat{\phi} (\boldsymbol{r},t)}{\partial t} = -
 \hat{\nabla}^2
\left( \hat{\nabla}^2 \hat{\phi} - \hat{\phi}^3+\hat{\phi} \right),
\label{cahn3}
\end{equation}
with
\begin{equation}
\begin{aligned}
\hat{\boldsymbol{\nabla}} & = \left( \frac{\partial}{\partial x'}, 
\frac{\partial}{\partial y'} - 
S (t) \frac{\partial}{\partial x'} \right),\\
\hat{\nabla}^2 & =  \left( \frac{\partial}{\partial x'} \right)^2 + 
\left( \frac{\partial}{\partial y'} - 
S (t) \frac{\partial}{\partial x'} 
\right)^2.
\end{aligned}
\end{equation}

After transformation, Eq.~(\ref{cahn3}) is formally identical to the 
Cahn-Hilliard equation without shear, which is solved
by the implicit spectral algorithm developed in Ref.~\cite{BBK1}.
Space is measured in units of the correlation length
$\xi$ (also the interface width) and time in units of 
$\xi^2 / \Gamma$.
This microscopic time scale represents
the typical time it takes to create a well-defined domain wall.
Periodic boundary conditions
are imposed in the deformed frame.
The single parameter of the simulation is then the shear
rate $\gamma$, which introduces a 
time scale $\gamma^{-1}$.
The choice of the parameters for the discretization
has been discussed in Ref.~\cite{BBK1} and 
the values $\Delta t = 0.5$, $\Delta x=\Delta y =0.5$ are
used throughout the simulation.
For each shear rate, the system size has been carefully checked
to be large enough so that the reported 
growth laws are unaffected by the boundaries. 
Since the growth is strongly anisotropic, 
a rectangular simulation box has been chosen with sizes up to 
$L_y = 512$ and $L_x = 8192$.
The shear rates investigated in this paper 
are $\gamma  = 0.04$, 0.02, 0.01, 0.005, 0.0025 and 0.00125.
This corresponds to a time scale $\gamma^{-1}$ 
in the range $[25,800]$. 
We wish to emphasize that the condition 
$\gamma^{-1} \gg \xi^2 / \Gamma$ has to be fulfilled, since 
we are interested in the scaling regime where 
well-defined domains coarsen.
This remark will become important for the interpretation
of the numerical results.
An alternative solution would be to apply the shear flow
after an initial transient, so that large domains have already
grown. 
Such initial conditions are discussed at the end of the paper.

\section{Zero shear case}

Although the $\gamma=0$ case has been 
extensively studied \cite{review_hohenberg,review_gunton,review_alan}, 
we briefly consider this well-known situation  
with three aims.
These results are presented to 
(i)  validate our numerical procedure,
(ii) present the quantities of interest and above all 
(iii)  make comparisons to the sheared case easier. 

\begin{figure}
\begin{center}
\begin{tabular}{ccc}
\psfig{file=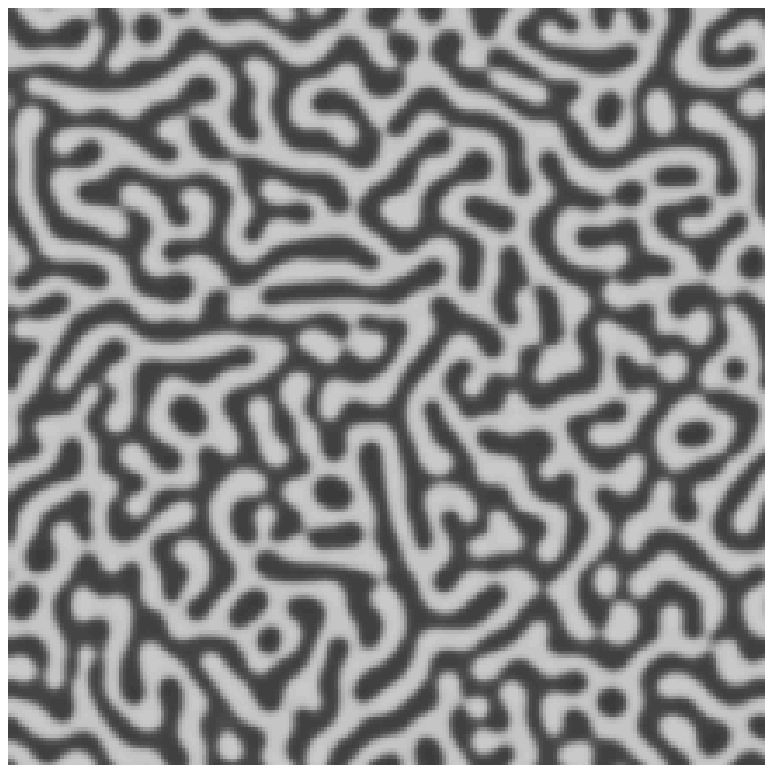,width=2.8cm,height=2.8cm} &
\psfig{file=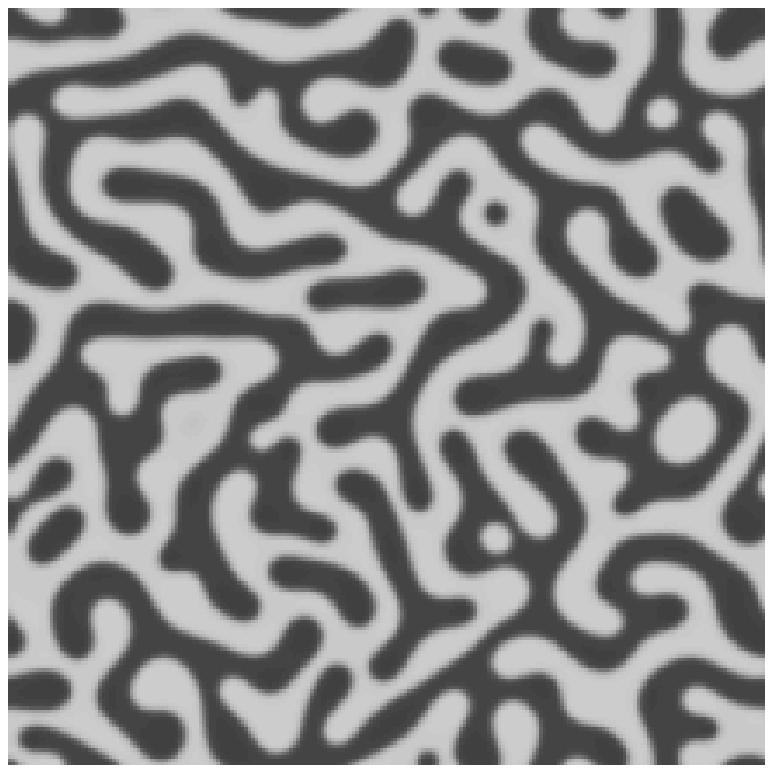,width=2.8cm,height=2.8cm} &
\psfig{file=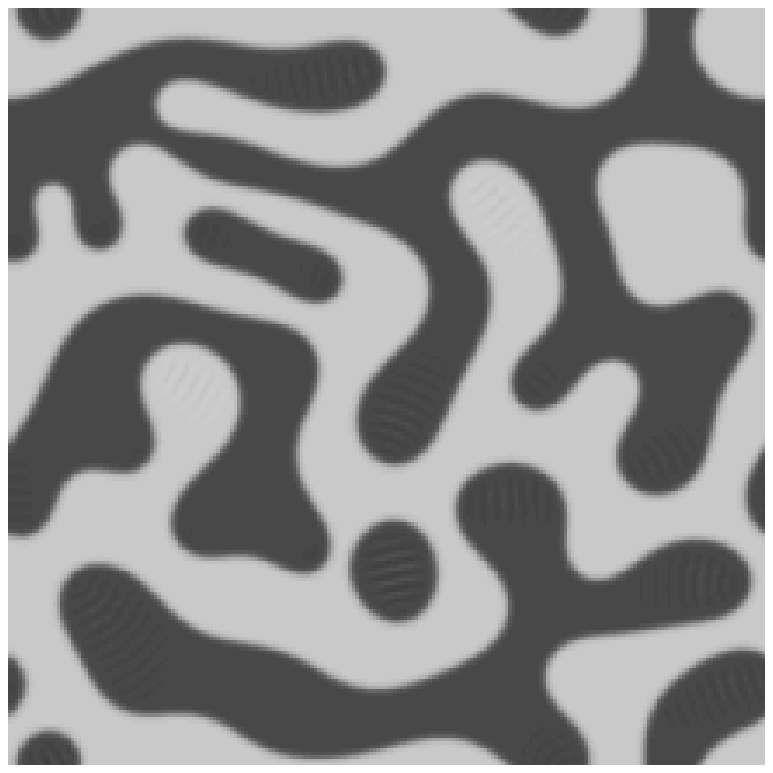,width=2.8cm,height=2.8cm} \\ 
\end{tabular}
\caption{Snapshots of size $256 \times 256$ of the unsheared 
spinodal decomposition at times $t=70$, 381 and 3761 (from left
to right).
Each color represents one phase of the mixture.}
\label{gamma=0gif}
\end{center}
\end{figure}

The domain growth takes place typically as in Fig.~\ref{gamma=0gif},
where an isotropic bicontinuous structure coarsens with time.
This coarsening process is well characterized by the two-point correlation
function defined by 
\begin{equation}
C(\boldsymbol{r},t) \equiv \frac{1}{V}
\int \upd^2 \boldsymbol{x} \langle \phi(\boldsymbol{x},t)
\phi(\boldsymbol{x+r},t) \rangle,
\label{2pointeq}
\end{equation}
which is nothing but the Fourier transform of the structure
factor, experimentally measured through light scattering experiments.
A typical two-point function is represented in Fig.~\ref{2point}
which shows the isotropy of this surface.
The average shape of the domains 
of Fig.~\ref{gamma=0gif} may be extracted  
from this plot by taking
the intersection of this surface with a horizontal plane
$z=constant$.
This allows to measure the time dependence
of the length scale $R_x(t)$ ($R_y(t)$) in direction $x$ ($y$).
Both length scales are represented in the inset of Fig.~\ref{scaling},
and have the expected power law behavior 
$R_x \simeq R_y 
\propto t^{1/3}$~\cite{review_hohenberg,review_gunton,review_alan}.

The dynamic scaling hypothesis is tested in the main frame of 
Fig~\ref{scaling}, where
the two-point function is circularly averaged and plotted as a
function of $|\boldsymbol{r}|/L(t)$.
This works perfectly well.

We are thus confident in our numerical setup, 
and we shall now address the question of the influence 
of the shear flow on the spinodal decomposition.

\begin{figure}
\begin{center}
\psfig{file=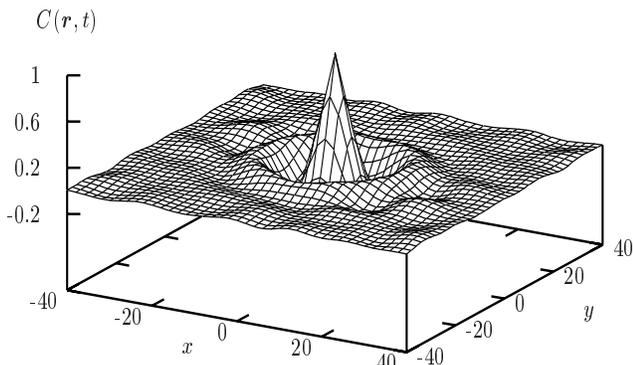,width=8.5cm,height=5.1cm}
\caption{Two-points correlation function, Eq.~(\ref{2pointeq}),
in the $\gamma=0$ case, 
at time $t=100$. 
\label{2point}}
\end{center}
\end{figure}

\begin{figure}
\begin{center}
\psfig{file=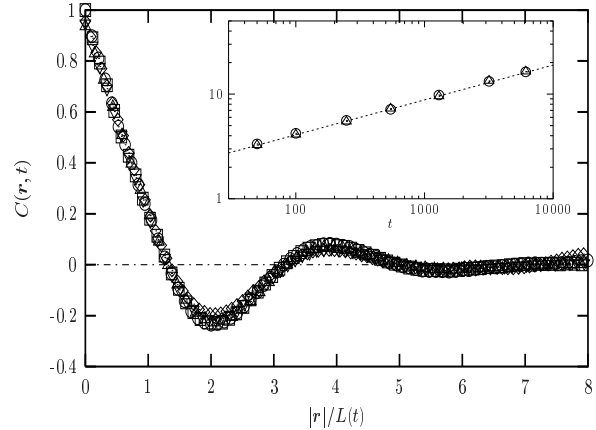,width=8.5cm,height=6.5cm}
\caption{MAIN: Circularly averaged two-point correlation 
as a function of the rescaled variable 
$|\boldsymbol{r}|/L(t)$ for different times in the range $[50,1000]$.
By construction, one has $C(L(t),t)=0.2$, the choice of 0.2 is indifferent.
INSET: Growth laws in the $x$ (circles) and $y$ (triangles) 
directions.
The dashed line is a fit to a power law $t^{1/3}$.}
\label{scaling}
\end{center}
\end{figure}

\section{Anisotropic growth of the domains under shear}

\subsection{Basic observations}

The time evolution of the domains 
when a shear flow is applied is followed in 
Fig~\ref{gammagif}. 
This evolution is the basic 
result of all previous numerical 
works~\cite{cogola,cates,shou,julia,zhang,yamamoto,cogola2,rothman,chan,qiu,padilla,ohta} 
and shows that the domain growth is essentially unaffected for times
$t \lesssim \gamma^{-1}$ since the first snapshot
is very similar to those in Fig.~\ref{gamma=0gif}.
At intermediate times, the domains begin to have an
anisotropic shape: an average direction of the domains is
clearly apparent.
This direction rotates and becomes more aligned with the flow
when the strain increases. 
At large times, $t \gg \gamma^{-1}$, 
domains are nearly aligned with the flow, and
have a strongly anisotropic shape.

These features are also clearly discernible in Fig.~\ref{2pointgamma},
where we have plotted the two-point correlation function for a
strain  $S(t)= 5$ at $\gamma= 0.01$.
The surface is clearly stretched in the flow direction, compare
with the Fig.~\ref{2point}.
In particular, it becomes impossible to perform
a circular average as in the unsheared case: isotropy is lost.

The average shape of the domains is again recorded
through  the intersection of the two-point correlation function
with a horizontal plane.
For definiteness, we take $z=0.5$, but the value
0.5 is unimportant.
It is found numerically that 
this intersection is very well represented by an ellipse.
Computing the parameters of this ellipse gives then 
access to two typical length scales, 
$R_{\parallel}(t)$ (large axis) and $R_{\perp}(t)$ (small axis), 
and to the mean orientation of the domains, $\theta(t)$ (angle
between the large axis and the $x$-direction).
We present the typical time evolution of the elliptic shape
of the domains in Fig.~\ref{ellipse}, where data
are also fitted to an elliptic form, in a very 
satisfactory way.

These three quantities depend both on the time $t$, and on
the shear rate $\gamma$ and
we successively study in the following subsections
the time evolution of $\theta(t)$ and of the length scales 
$R_{\parallel}(t)$ and $R_{\perp}(t)$.

\begin{figure}
\begin{center}
\psfig{file=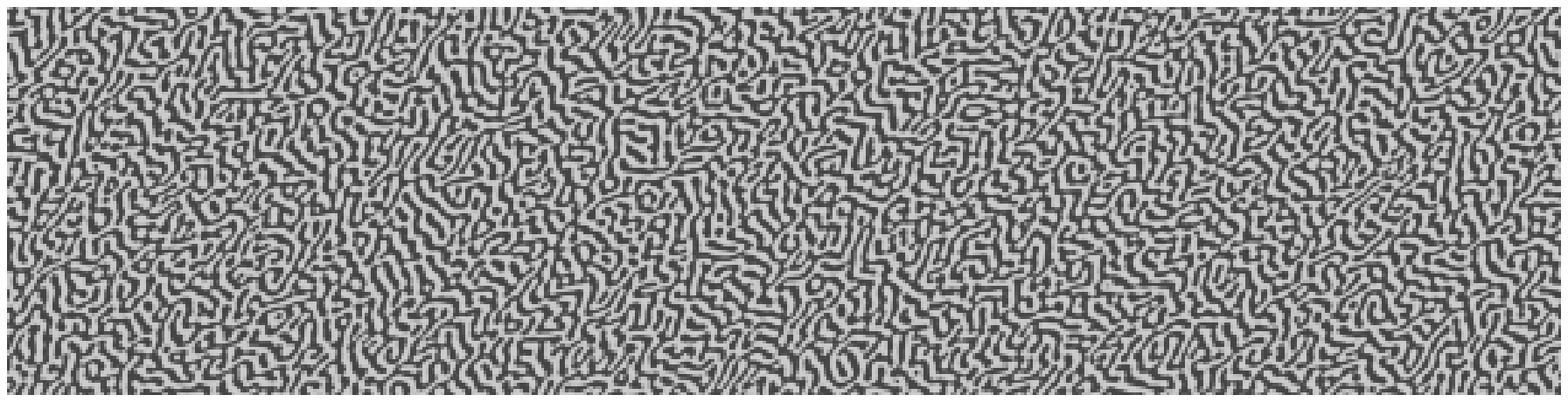,height=2.2cm,width=8.8cm} \\
\psfig{file=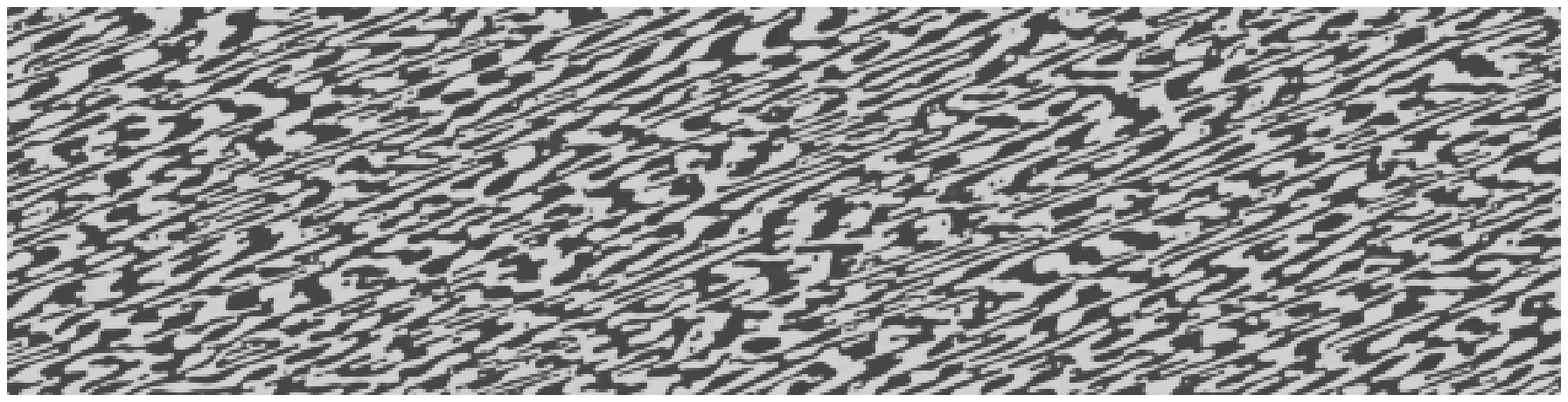,height=2.2cm,width=8.8cm} \\
\psfig{file=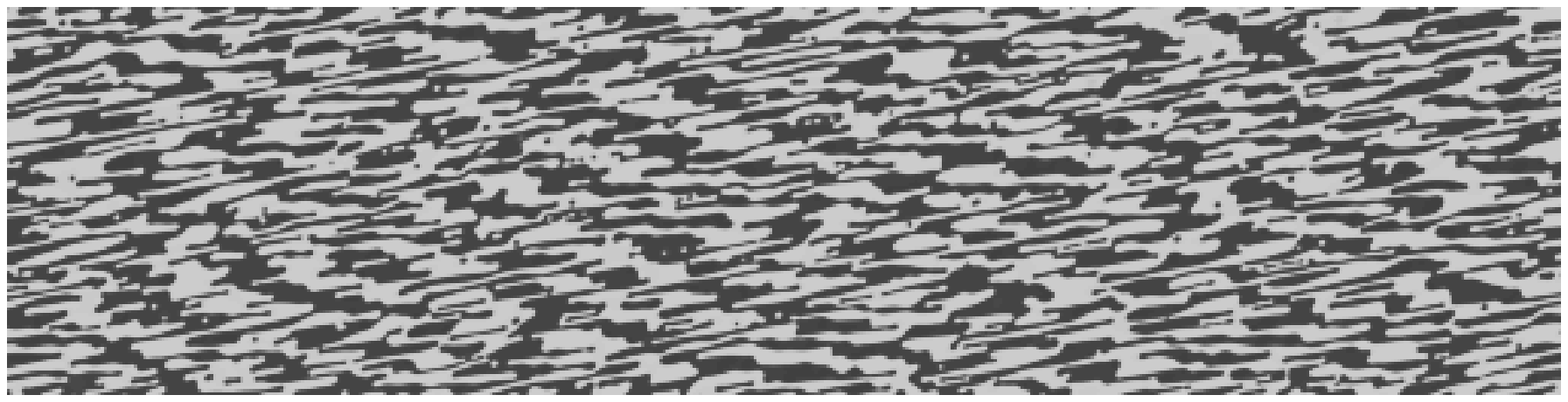,height=2.2cm,width=8.8cm} \\
\psfig{file=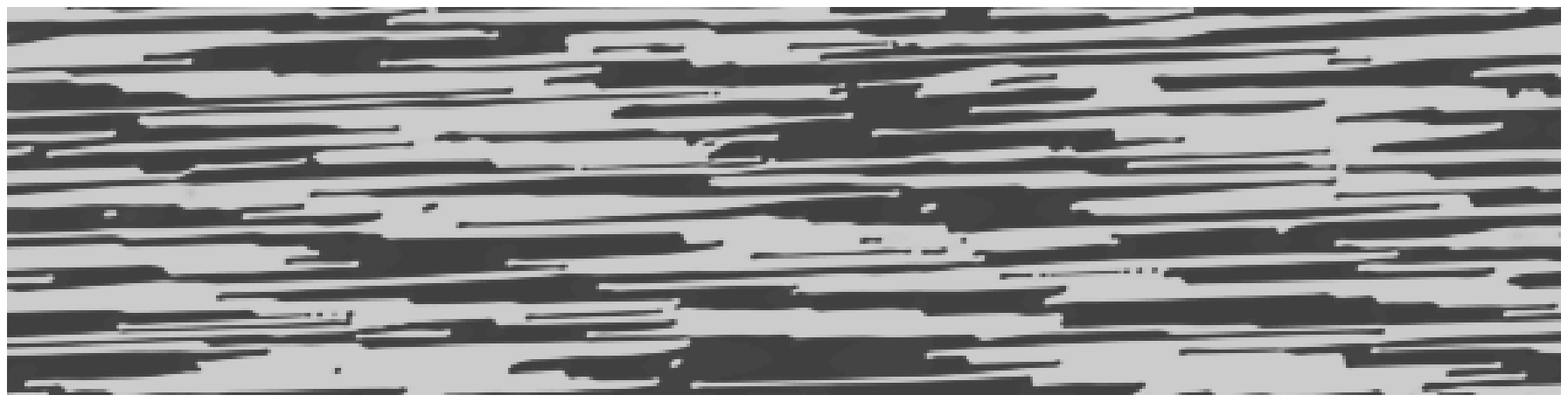,height=2.2cm,width=8.8cm}
\caption{Snapshots of size 
$L_y = 512$ and $L_x = 2048$ (parts of a $512 \times 4096$ system) 
for a shear rate $\gamma = 0.01$ and strains 
$S(t)=1$, 5, 10 and 50 (from top to bottom).
Each color represents one phase of the mixture.}
\label{gammagif}
\end{center}
\end{figure}

\begin{figure}
\begin{center}
\psfig{file=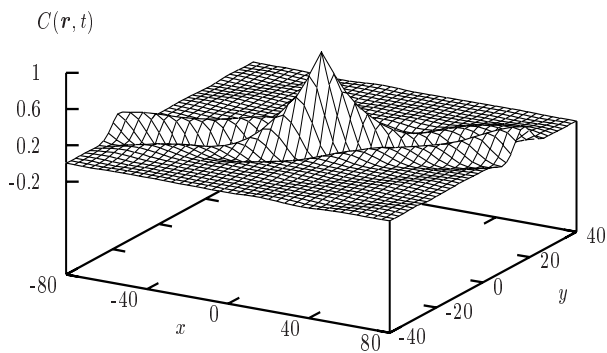,width=8.5cm,height=5.25cm}
\caption{Two-point correlation function for $\gamma=0.01$ and
$S(t) = 5$. The surface is stretched in the $x$-direction.
Note, in particular, that the $x$ and $y$ ranges 
are different in this figure.    
\label{2pointgamma}}
\end{center}
\end{figure} 

Before performing this analysis, a remark has to be made
about the identification of the relevant length scales.
It is quite clear from the above analysis that 3 parameters
are needed to fully characterize the growing structure under shear.
This feature was already noted in experiments~\cite{beysens0,beysens1}.
In some of the previous numerical works, only two parameters were 
studied, namely the length scales in the $x$ and $y$ directions. 
Although these length scales should well represent
the structure at long times, it is physically preferable
to study instead the domain size in the directions defined by the angle
$\theta(t)$.
In Ref.~\cite{julia}, the direction of the domains'
shape was recorded, but there was no attempt at a quantitative analysis
of its behavior.
We note last that this quantitative analysis of the
domain morphology in two dimensions
naturally arises from the analytical work of Ref.~\cite{brca},
in the case of a non-conserved order parameter.

\begin{figure}
\begin{center}
\psfig{file=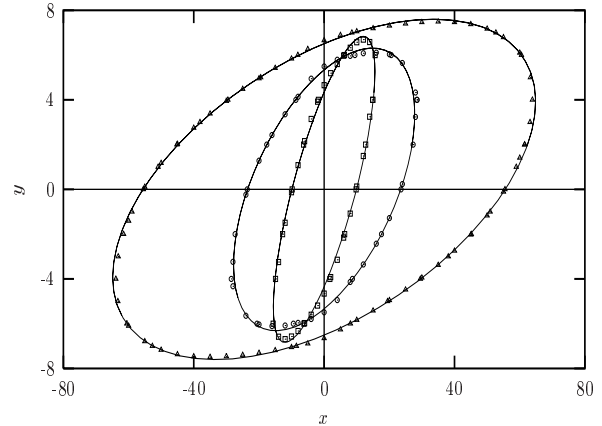,width=8.5cm,height=6.5cm}
\caption{Time evolution of the intersection 
of the two-point correlation
with a horizontal plane, for strains $S(t) = 5$ (squares), 
10 (circles) and 20 (triangles) and shear rate $\gamma=0.01$.
The points are the data, while the lines are fits to an elliptic
shape.
Note, in particular, that the $x$ and $y$ ranges 
are different in this figure.   
\label{ellipse}}
\end{center}
\end{figure} 

\subsection{The mean orientation $\boldsymbol{\theta(t)}$}

The first effect of the shear flow is to give
the domains an anisotropic shape and hence to
create a preferred direction in the system. 
Physically, it can be expected that for a given shear rate
$\gamma$, the mean orientation
of the domains is 
inversely proportional to the strain $S(t)$, as would
be the case for a rigid rod advected by the shear flow:
\begin{equation}
\theta(t) \simeq \frac{\theta_0(\gamma)}{\gamma t}.
\label{strain}
\end{equation}
This relation is tested in Fig.~\ref{theta}, 
where $\theta_0(\gamma)$ is
used as an adjustable parameter.
This figure shows that the relation (\ref{strain}) is well 
satisfied in the whole range of shear rates investigated.

The parameter $\theta_0$ is found to be a slowly increasing 
function of the shear rate.
One finds $\theta_0(0.00125) \simeq 0.975$ and $\theta_0(0.04) \simeq 2.0$.
By definition, this angle corresponds to
the mean orientation 
of the domains when the strain is 1, 
$\theta_0 \equiv \theta(t=\gamma^{-1})$, and its variation
may be understood by the following argument.
For times $t \le \gamma^{-1}$, the domain growth has been mainly
unaffected by the flow and thus the larger the shear rate, the smaller
the domains at time $t \sim \gamma^{-1}$.
Since large domains are more easily deformed than small ones
[because of the surface tension], it is
expected that at strain $S(t) \sim 1$,
large domains (small $\gamma$) 
are more deformed than small ones (large $\gamma$).
Hence, the smaller the shear rate, the smaller $\theta_0$.

\begin{figure}
\begin{center}
\psfig{file=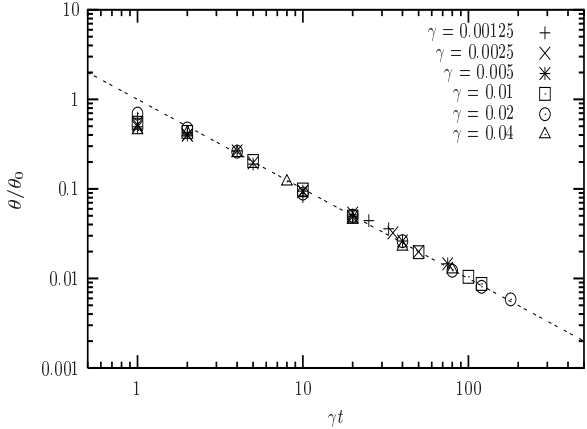,width=8.5cm,height=6.5cm} 
\caption{Mean orientation of the domains as a function of 
the strain $S(t) = \gamma t$. The straight line is $1/\gamma t$.}
\label{theta}
\end{center}
\end{figure}

\subsection{Growth laws}

We turn now to the time evolution of the two length scales
$R_\perp (t)$ and $R_\parallel (t)$.
These quantities are studied for a very broad range of shear rates
from $\gamma = 0.04$ 
($\gamma^{-1} = 25$) to $\gamma = 0.00125$ 
($\gamma^{-1} = 800$).
All our results are summarized in 
Fig.~\ref{growth}, where $R_\parallel$ and $R_\perp$ are 
represented for each value of the shear rate as functions of the strain.
We obtain numerically that the growth laws are well
represented at large strains by the following algebraic forms, 
\begin{equation}
R_{\parallel}(t) \simeq R_{\parallel 0} \cdot 
(\gamma t)^{\alpha_{\parallel}}, \quad
R_{\perp}(t) \simeq R_{\perp 0} \cdot (\gamma t)^{\alpha_{\perp}},
\label{exponent}
\end{equation}
which define the growth exponents $\alpha_\parallel$ and $\alpha_\perp$.
This was first obtained in a simulation by
Padilla and Toxvaerd~\cite{padilla} and subsequently in similar systems in
Refs.~\cite{qiu2,cogola,cogola2,brca}.

In Fig.~\ref{growth}, we have fitted our data using 
the forms (\ref{exponent}).
These fits deserve some comments. 
The growth law for $R_{\parallel}$ becomes 
a nice straight line in a log-log plot at large strains 
$S(t) \gtrsim 10$. 
Only the data points for such large strains have been used to 
compute the exponent $\alpha_\parallel$. 
Concerning the time behavior of $R_\perp (t)$, two different fits
have been tested.
First, a horizontal line corresponding to $\alpha_\perp = 0$ has been
compared to the data.
Second, following the available analytical 
results~\cite{cogola,rabr,brca},
we also tried  to fit the data with the ansatz
\begin{equation}
\alpha_\perp = \alpha_\parallel - 1.
\label{ansatz}
\end{equation}
Note that both fits are equivalent when $\alpha_\parallel =1$, which
is nearly the case for the two smallest shear rates.

Several comments are in order.

1 - The algebraic fits are clearly a very good representation of the data.
Some logarithmic corrections that come out from 
the analytical study of the $O(n)$ model and of the non-conserved
case may be present, but we do not expect
to be able to determine them numerically.

2 - Although the relation (\ref{ansatz}) reasonably accounts
for the data, the value $\alpha_\perp = 0$ is also 
possible and works well for {\it all} the shear rates investigated.
This means that it could be possible to rescale all
the curves for $R_\perp (t)$ by plotting 
$R_\perp(t) / R_{\perp 0}$ as a function of the strain for 
different shear rates, using $R_{\perp 0}$ as a fitting parameter.
This parameter is found to be a decreasing function of the shear rate,
which means that the smaller the shear rate, the wider
the domains. 
This rescaling is performed in Fig.~\ref{rescperp}, and the data 
are indeed compatible with this hypothesis.
Let us add the remark that $R_\perp(t)$ varies (at most)
by a factor 2 in all the simulations. 
This strongly supports the hypothesis that there is in fact no
growth in the perpendicular direction.
Of course, once again, logarithmic corrections can not be 
numerically dismissed.

3 - We do not see any evidence for the oscillations reported
in Refs.~\cite{cogola,cogola2}.
We do not have a clear explanation for that, but
an hypothesis is that these oscillations are a preasymptotic
artifact of the measure process itself.
The definition of the length scales used in Ref.~\cite{cogola}
leads in some cases to
a ratio $R_x(t) / R_y(t) \simeq 10$ 
for a strain $S(t) \simeq 1$~\cite{cogola}. 
For this strain, the domains are still nearly circular and
in our simulations, this ratio is never above the value 1.5.
However, this argument is made weaker by the observation
that Qiu {\it et al.}~\cite{qiu2}
have used the same measurement procedure
as Corberi {\it et al.}~\cite{cogola}, at a lower shear 
rate $\gamma=0.01$ and do not observe any oscillations.

4 - An important point is the fact that the exponents
apparently depend on the value of the shear rate.
More precisely, we find that $\alpha_\parallel$ decreases from
the value $\alpha_\parallel \simeq 1.35$ at $\gamma = 0.04$
 to $\alpha_\parallel \simeq 1.0$ at $\gamma = 0.00125$. 
This means 
that we are in fact measuring {\it effective exponents} and 
that the true asymptotic behavior has not been reached in some 
of the cases studied here.

The problem is then to determine which exponent is the right one!
As emphasized in the Introduction,
well-defined interfaces exist only if $\gamma \ll 1$.
This indicates that the `true' asymptotic behavior
is reached for the smallest shear rates investigated, and favors
the value $\alpha_\parallel = 1$ found for
$\gamma = 0.0025$ and 0.00125.
Together with the behavior of $\alpha_\perp$ and the observation
that the domains are wider at lower shear rates, we are led
to the conclusion that the growth exponents are given by
\begin{equation}
\alpha_\parallel = 1,\quad \alpha_\perp = 0.
\end{equation}
The analysis of dynamic scaling properties 
in the following section will reinforce 
this conclusion.

\begin{figure}
\begin{center}
\psfig{file=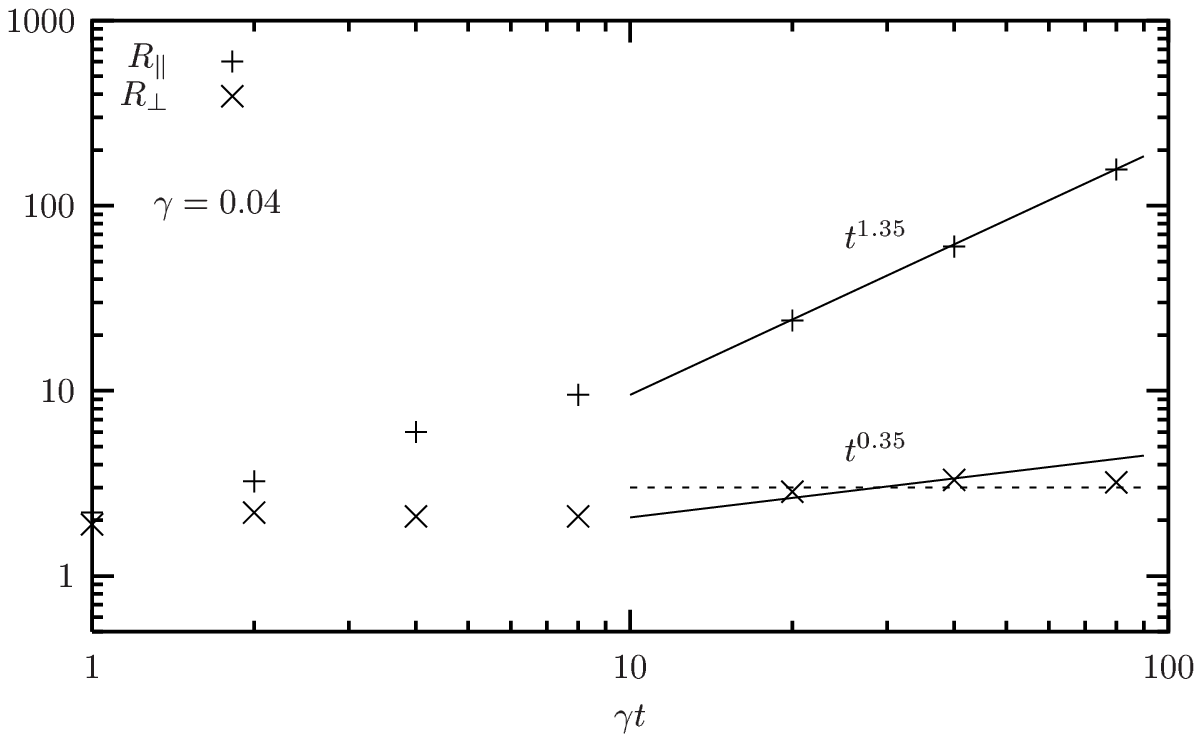,width=8.5cm,height=6.5cm} \\
\psfig{file=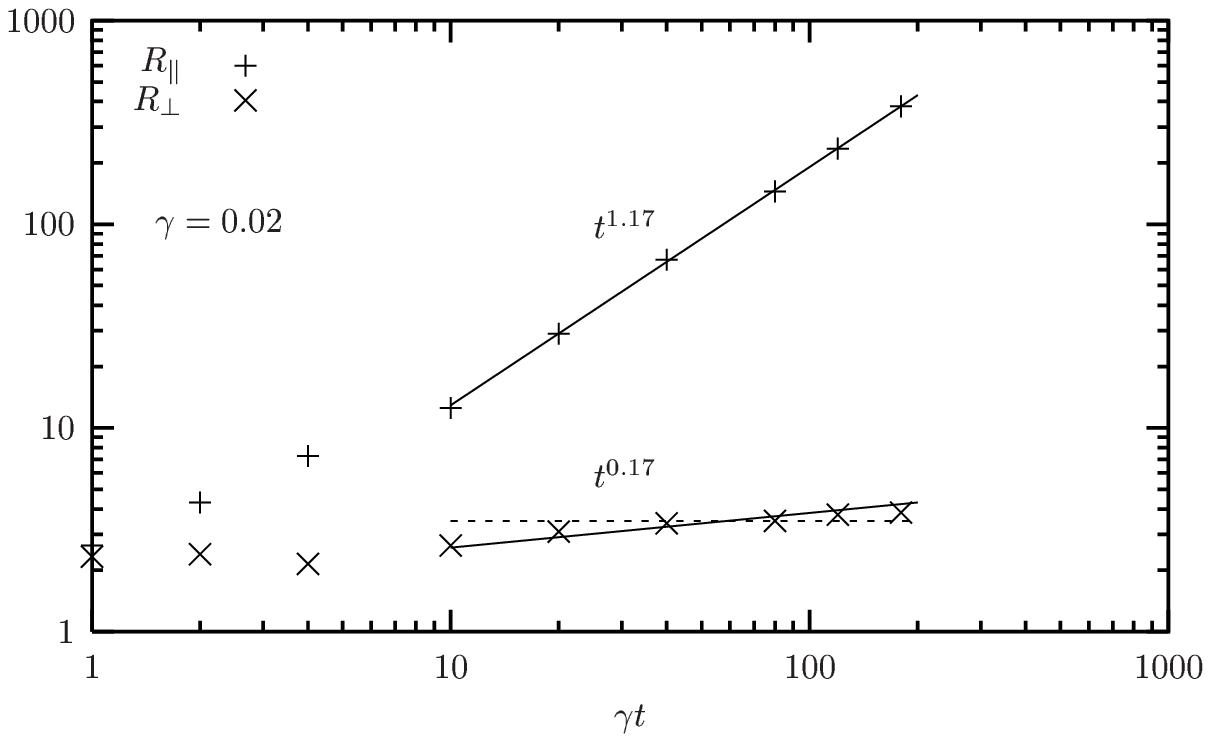,width=8.5cm,height=6.5cm} \\
\psfig{file=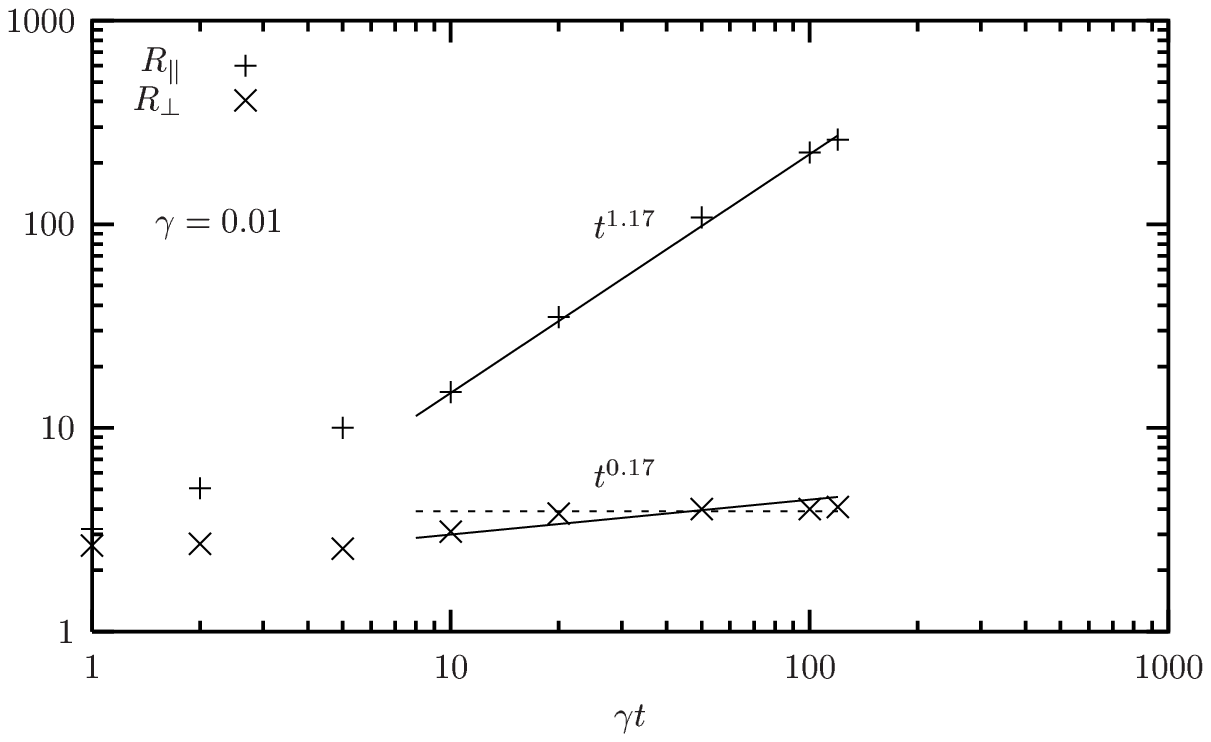,width=8.5cm,height=6.5cm} \\
\psfig{file=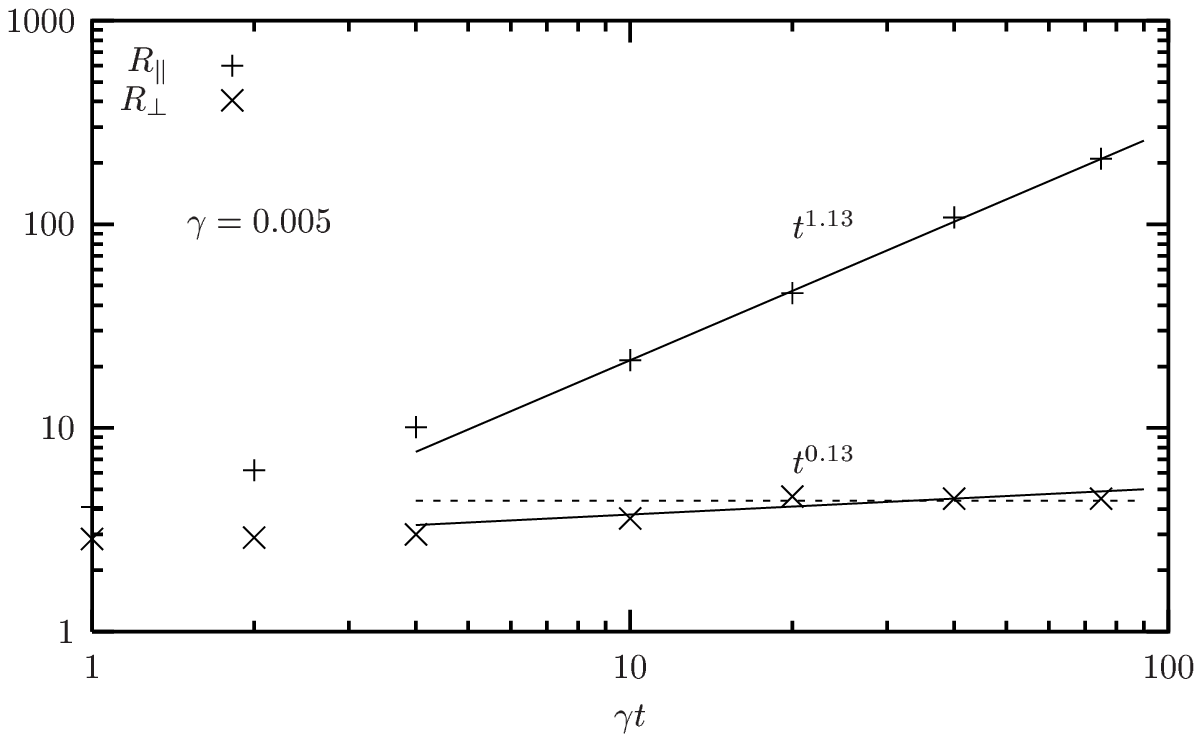,width=8.5cm,height=6.5cm} \\
\psfig{file=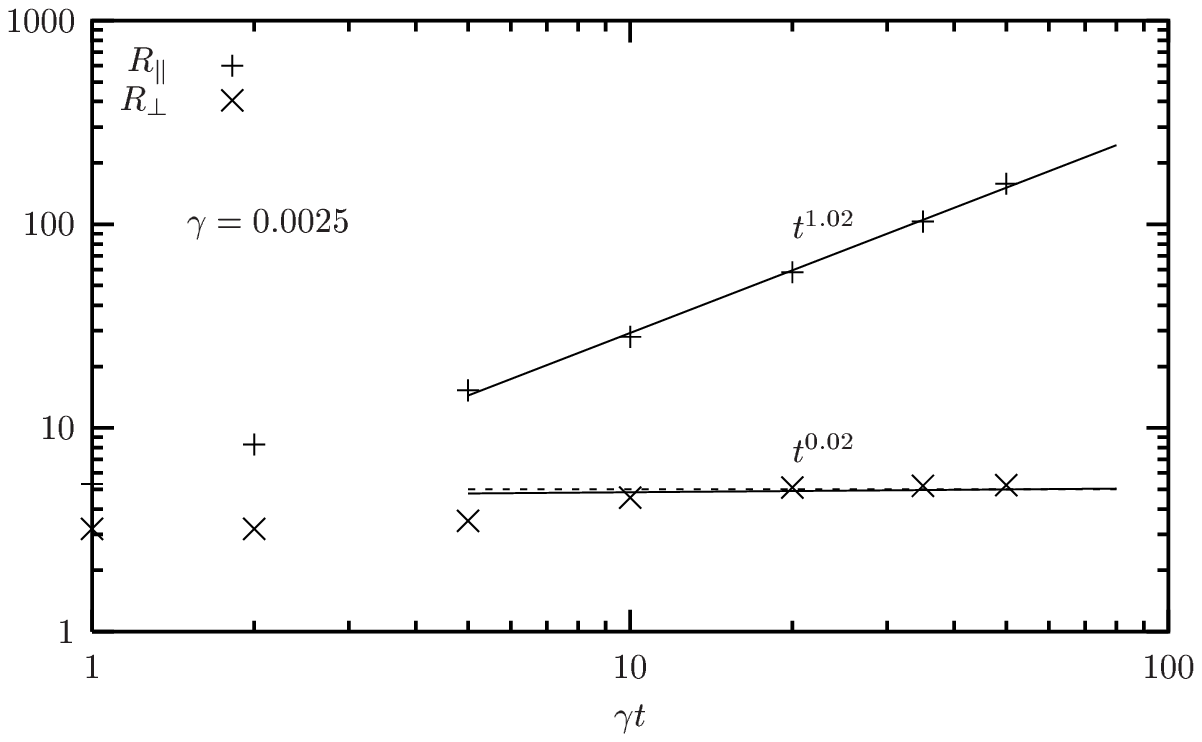,width=8.5cm,height=6.5cm} \\
\psfig{file=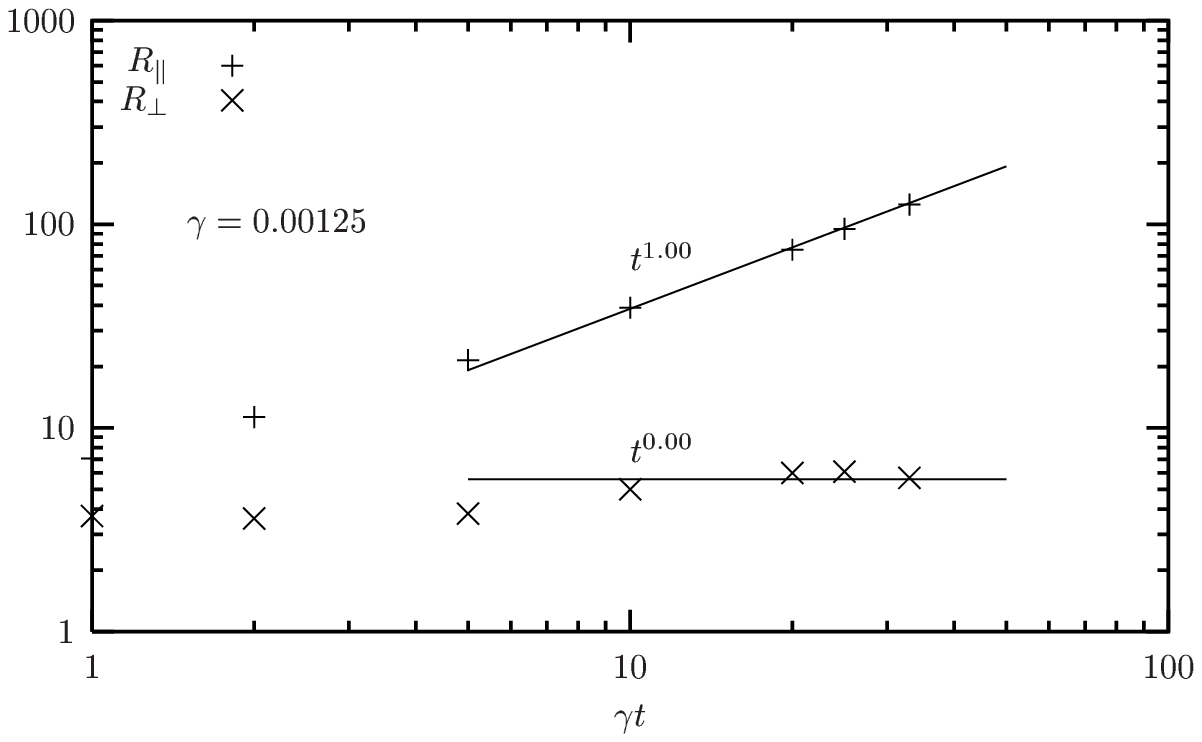,width=8.5cm,height=6.5cm} \\
\caption{The two length scales
$R_\perp (t)$ and $R_\parallel (t)$ are represented
as a function of the strain.
Each figure is labeled by the corresponding shear
rate.
The symbols are the data, and the full lines are algebraic fits 
with the exponents indicated near each fit.
For the strongest shear rates, a horizontal dashed line
has been added as a fit to $\alpha_\perp =0$.}
\label{growth}
\end{center}
\end{figure}   

\begin{figure}
\begin{center}
\psfig{file=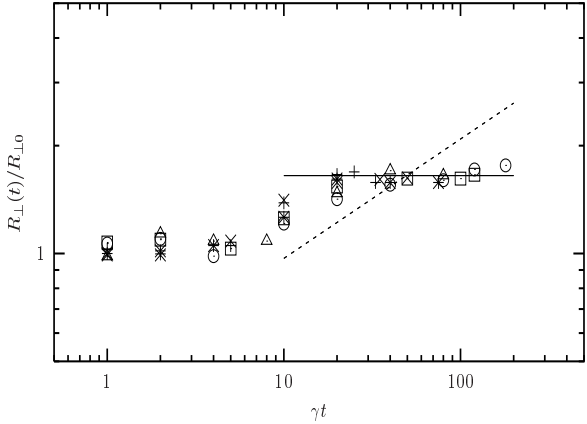,width=8.5cm,height=6.5cm} 
\caption{Test of the relation (\ref{exponent}) for the small
axis of the ellipse. 
The full line represents the case $\alpha_\perp = 0$, whereas
the dashed line is for $\alpha_\perp = 1/3$.
The symbols are the same as in Fig.~\ref{theta}.
The latter is clearly inconsistent with the numerical measurements
for all the shear rates.}
\label{rescperp}
\end{center}
\end{figure}

\section{Dynamic scaling}

The last question we wish to address is the problem of dynamic scaling.
We have recalled in Fig.~\ref{scaling} (where $\gamma =0$) 
that the two-point correlation function
has the property that it can be rescaled in the
form of a single variable function
\begin{equation}
C(\boldsymbol{r},t) = {\cal C} \left( \frac{|\boldsymbol{r}|}{L(t)} 
\right).
\label{scaling0}
\end{equation}
This indicates that $L(t)$ in the only relevant length
scale in the asymptotic regime characterized by $\xi \ll L(t)$.
This behavior is the basis of a scaling argument
which, in the unsheared case, allows an elegant derivation
of the growth laws~\cite{alan_rg}.

In the shear flow, 
there are two relevant length scales and 
the scaling (\ref{scaling0}) can thus no longer be true.
Two different generalizations have been suggested by
analytical works.
The solution of the $O(n)$ model predicts the scaling 
form~\cite{rabr}
\begin{equation}
C(\boldsymbol{r},t) = {\cal C} \left( \frac{x}{R_x(t)},\frac{y}{R_y(t)}
\right),
\label{scaling1}
\end{equation}
where $R_x$ and $R_y$ are the typical sizes in the directions
$x$ and $y$ respectively.
A different scaling is expected from the non-conserved order parameter case,
namely \cite{brca}
\begin{equation}
C(\boldsymbol{r},t) = {\cal C} \left( \frac{r_{\parallel}}{R_{\parallel}(t)},
\frac{r_{\perp}}{R_{\perp}(t)} \right),
\label{scaling2}
\end{equation}
where the subscripts refer to the rotating frame described 
previously.
Corberi {\it et al.}~\cite{cogola} have used the form (\ref{scaling1})
as a starting point
to generalize the argument of Bray~\cite{alan_rg} to the sheared
case. It is thus important to see if 
this scaling behavior is detected in the simulation.
Let us note that the tilt angle
$\theta(t)$ is very small in the long time regime 
we are interested in.
Then, one might ask if the difference between 
the two forms (\ref{scaling1})-(\ref{scaling2}) proposed 
above is relevant.
Since the domains are very elongated in the $x$-direction, then
even with a small angle there might be differences 
between the `parallel' and the $x$-directions.
Concerning the $y$ and the `perpendicular' directions, the support
of the correlation function in these directions is very small, 
so that differences between the two are indeed  not observable.

\begin{figure}
\begin{center}
\psfig{file=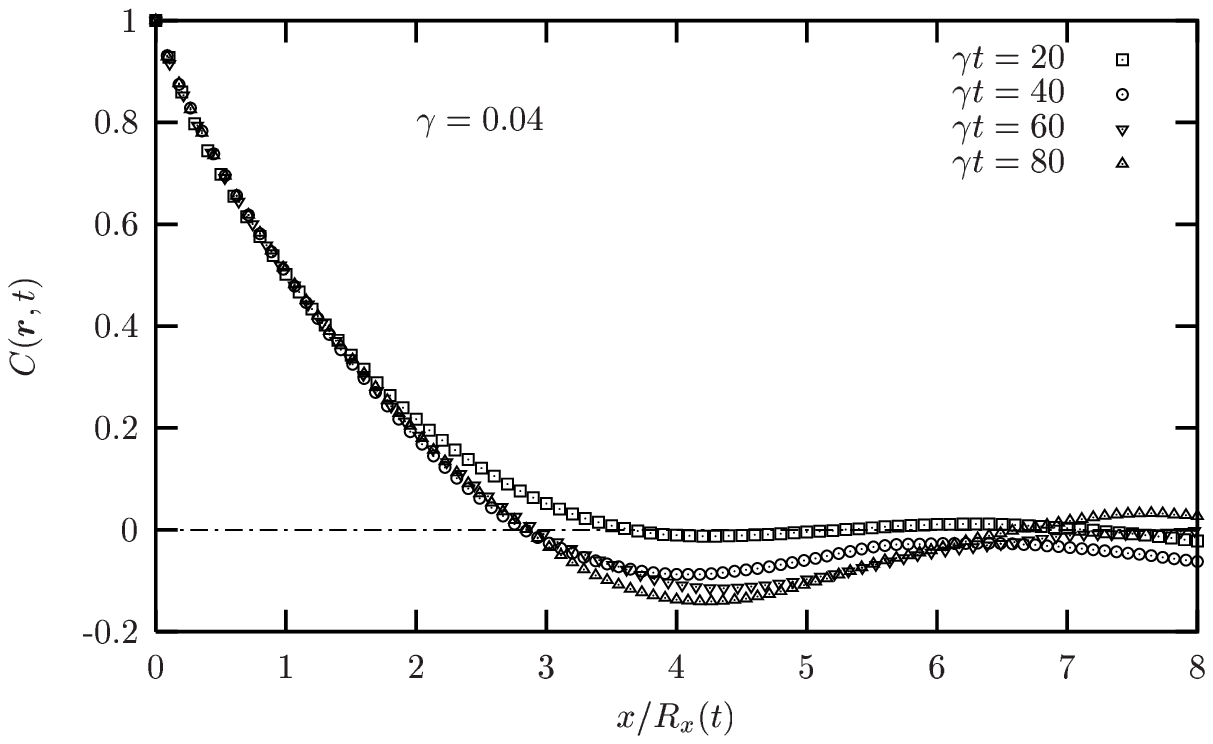,width=8.5cm,height=6.5cm} \\
\psfig{file=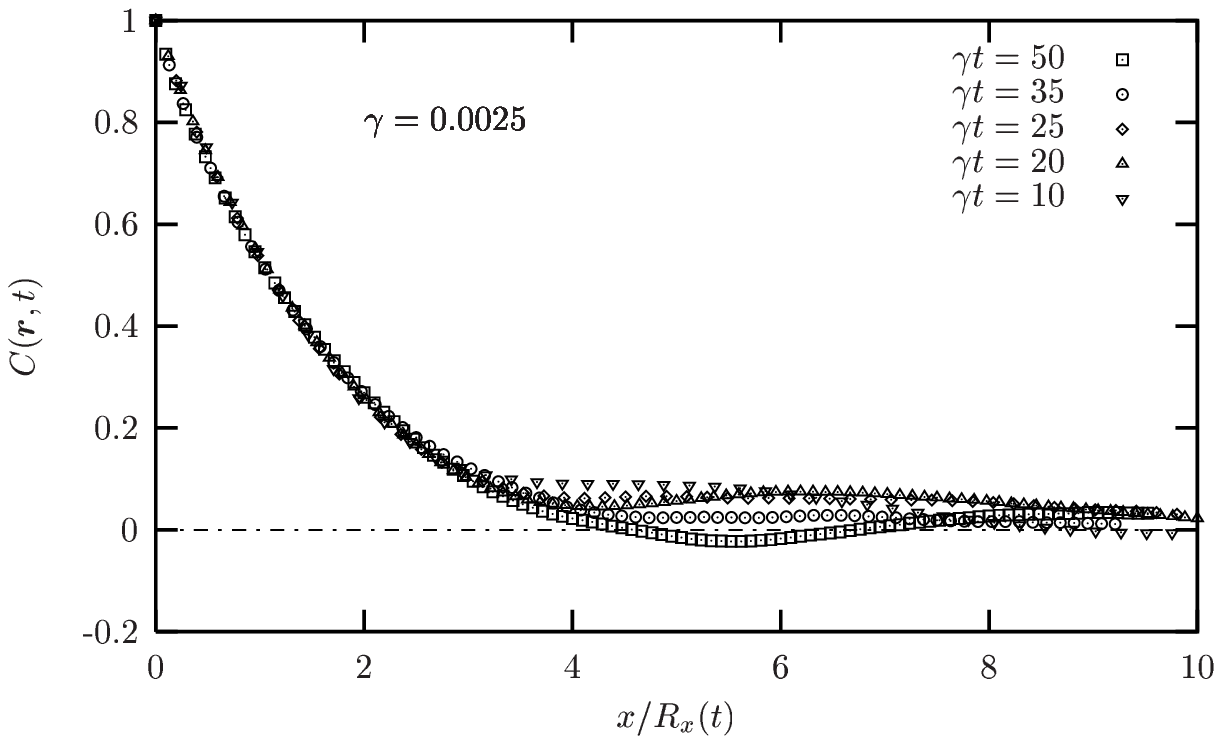,width=8.5cm,height=6.5cm} \\
\caption{Test of dynamic scaling in the $x$-direction for 
$\gamma=0.04$ and 0.0025. 
In both cases, the collapse is not satisfying.}
\label{resc1}
\end{center}
\end{figure}

We now present our numerical results.
In Fig.~\ref{resc1}, we show for two different shear rates 
the attempt at the rescaling of the two-point function in the $x$-direction.
In each case, we have considered a time window where the hypothetic
scaling might hold ({\it i.e.} the growth law has reached
its algebraic asymptotic form, $S(t) \gtrsim 10$).
We have also chosen $R_x(t)$ in order to get the best collapse of the 
data. Clearly, the scaling (\ref{scaling1}) does not work, in
both cases of a high and a low shear rate.

We investigate the possibility for the second scaling form to
hold in the direction defined by the tilt angle $\theta(t)$
in Fig.~\ref{resc2}.
There is here a clear qualitative difference between the shear rate
$\gamma=0.0025$ and $\gamma=0.04$, since the collapse
of the data is excellent for the small shear rate, whereas
there is a clear systematic evolution for the highest shear rate.

In our opinion, 
these results are a very good indication that
the scaling form (\ref{scaling1}) does not describe
the asymptotic behavior of the Cahn-Hilliard equation.
Moreover, they show that a self-similar asymptotic regime
has been reached for the lowest shear rates investigated,
characterized by the scaling 
(\ref{scaling2}) of the two-point function.

This is also confirmed by the inspection
of the scaling properties in the `parallel' direction
(which is equivalent to the $y$-direction).
In Fig.~\ref{resc3}, we show the results for this direction.
Once again, the collapse is very good for the low shear rate 
$\gamma=0.0025$, whereas it is clearly not satisfying 
for a higher shear rate $\gamma=0.04$.

\begin{figure}
\begin{center}
\psfig{file=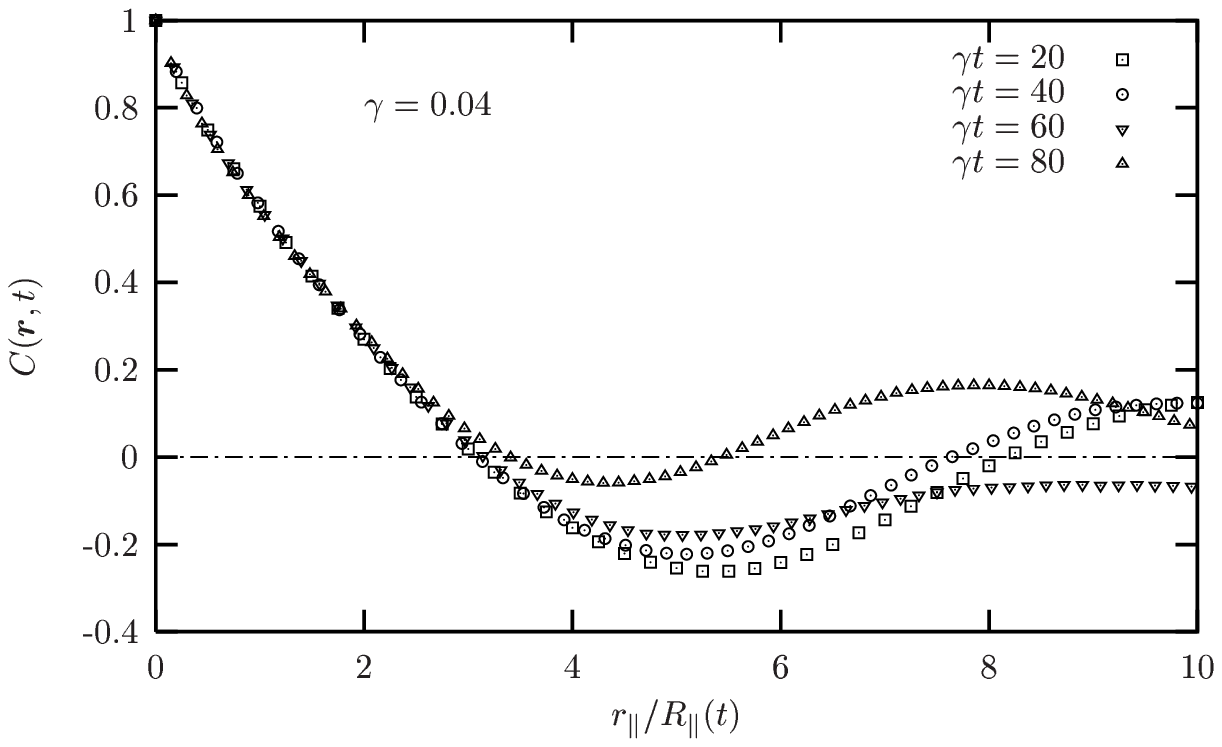,width=8.5cm,height=6.5cm} \\
\psfig{file=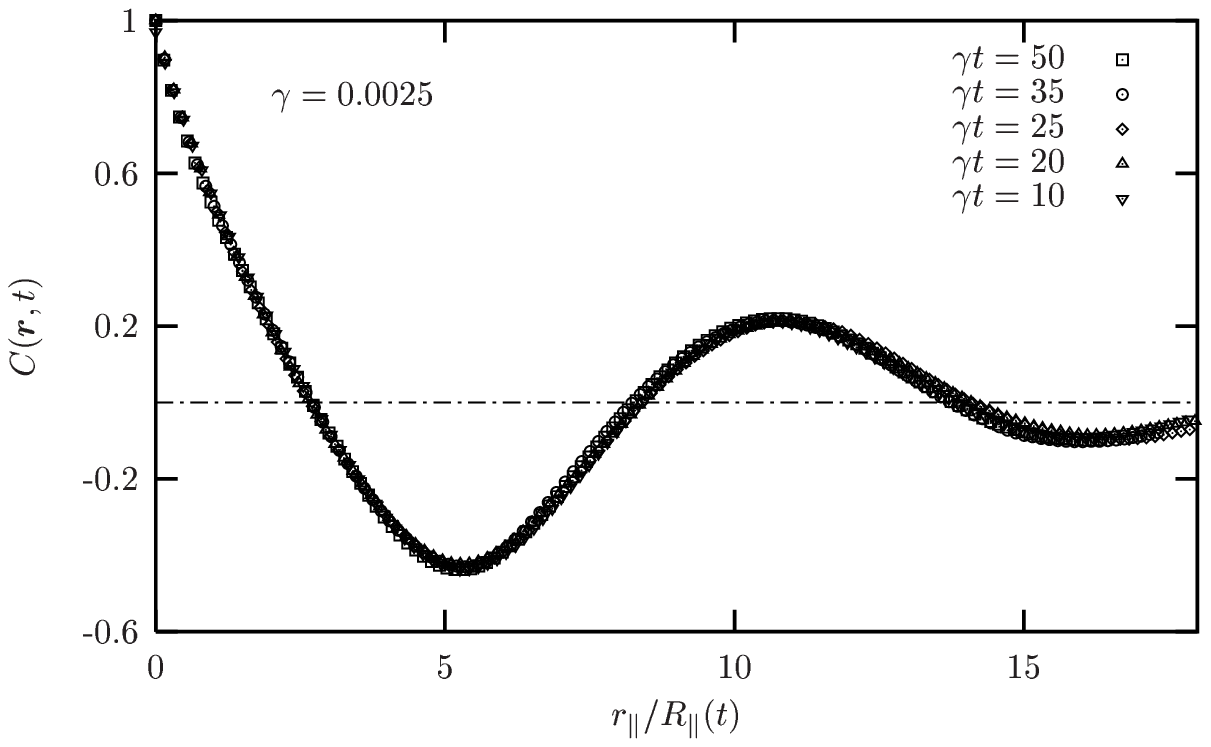,width=8.5cm,height=6.5cm} \\
\caption{Test of dynamic scaling in the `parallel' direction for 
$\gamma=0.04$ and 0.0025.
The collapse is very good for 0.0025 only, while there is a systematic
evolution for 0.04.}
\label{resc2}
\end{center}
\end{figure}

\begin{figure}
\begin{center}
\psfig{file=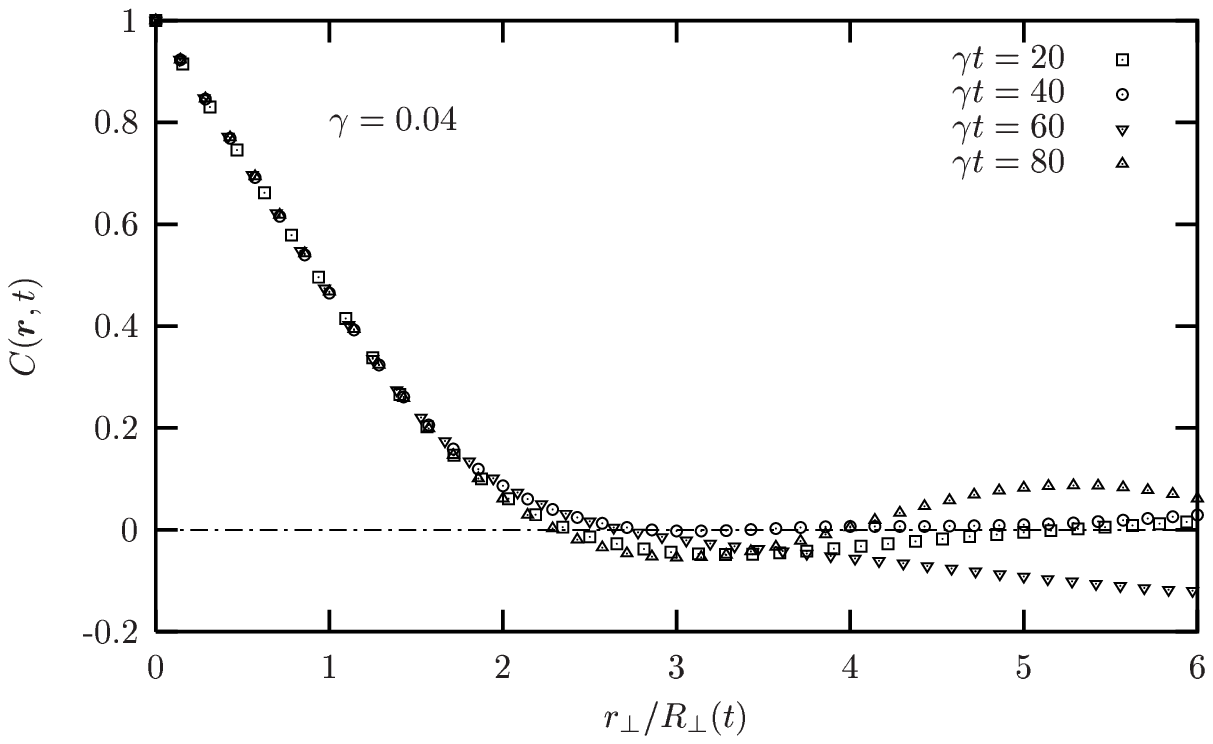,width=8.5cm,height=6.5cm} \\
\psfig{file=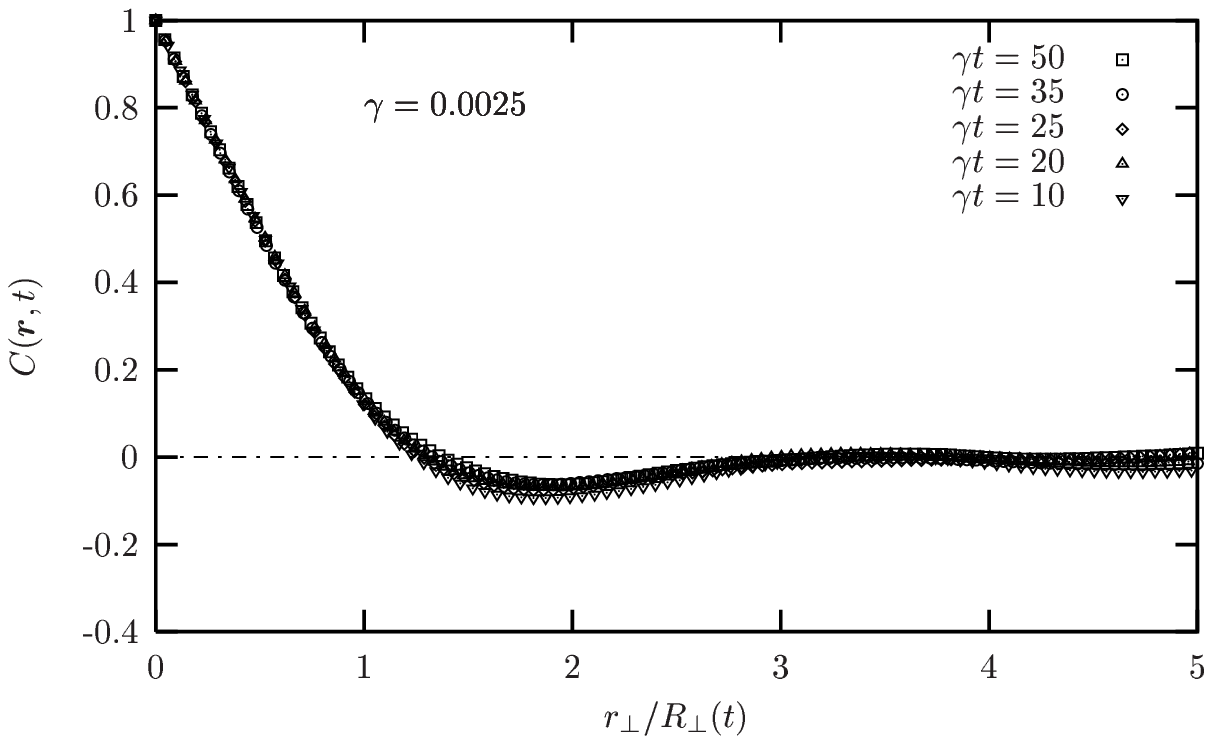,width=8.5cm,height=6.5cm} \\
\caption{Test of dynamic scaling in the parallel (or equivalently $y$)
direction for $\gamma=0.04$ and 0.0025. The collapse is very 
good for the lowest shear rate.}
\label{resc3}
\end{center}
\end{figure} 

\section{Discussion}

We have investigated in this paper the ordering kinetics
of a binary mixture quenched below its spinodal line 
in a homogeneous shear flow, through the numerical solution 
of the Cahn-Hilliard equation in 2 spatial dimensions.

We have found that the bicontinuous coarsening structure
is well described by three parameters. 
Since the average shape of the domains is elliptic, 
it is sufficient to give the lengths $R_\perp (t)$ and
$R_\parallel (t)$ of the two principal 
axis of the ellipse, and the orientation of the large axis 
with respect to the flow, $\theta (t)$.
We have measured the time evolution of these three
quantities, defined from the two-point correlation
function, and all our results may be summarized by the 
following relations, which
have been shown to hold
in the asymptotic regime of large times
\begin{equation}
\begin{aligned}
R_{\parallel}(t) & \simeq \gamma t ,\\
R_{\perp} (t) & \simeq constant, \\
\theta(t) & \simeq \frac{1}{\gamma t}, \\
C(\boldsymbol{r},t) & \simeq {\cal C} \left( 
\frac{r_{\parallel}}{R_{\parallel}(t)},
\frac{r_{\perp}}{R_{\perp}(t)} \right).
\label{regime}
\end{aligned}
\end{equation}
In particular, this asymptotic regime could not be 
reached for the largest
shear rates we have investigated.
Moreover,
our attempts at detecting (in the cases
$\gamma=0.04$ and 0.02) a crossover from a preasymptotic
regime with $\alpha_\parallel > 1$ and $\alpha_\perp > 0$ towards
the regime (\ref{regime}) were unsuccessful, because this 
would require too large a system size $L_x$.
These results also demonstrate that the domain growth
does not stop under the shear flow, in the case where
hydrodynamics is neglected
and answer then questions (1)-(4) of the introduction.

We now compare our results with previous ones.
It is interesting to note that, 
even though this simulation was not intended to be able to reproduce
experiments since hydrodynamics has been neglected, our results
reproduce quantitatively
the experiments of Chan {\it et al.}~\cite{beysens1,beysens2} on
a mixture of water and isobutyric acid
and of Qiu {\it et al.}~\cite{qiu} on a polymer blend of polystyrene and 
poly-(vinyl methyl ether),
for the three quantities
describing the morphology of the domains. 
These experiments are probably done in a regime where
hydrodynamic effects are negligible, since other
experiments by Hashimoto and collaborators~\cite{string}
report a saturation  of the domains to a $\gamma$-dependent size.
This stationary steady state has been termed
`string phase' and has also been
observed by Hobbie {\it et al.}~\cite{hobbie} in a strong shear
rate regime.

Our numerical findings can be compared with previous 
works neglecting hydrodynamics, and
our results for the growth
laws are similar to those found by Qiu 
{\it et al.}~\cite{qiu} and Corberi {\it et al.}~\cite{cogola}.
Note that
in this last reference~\cite{cogola}, the
growth laws are compatible with the regime (\ref{regime}), 
although they are compared with the behavior $R_x(t) \sim t^{4/3}$ and
$R_y(t) \sim t^{1/3}$.
The authors admit however that the latter regime
was not reached within the numerical time-window.
Earlier simulations were not able
to be very quantitative~\cite{rothman,chan,padilla,ohta}, although
Padilla and Toxvaerd~\cite{padilla} have suggested that the 
anisotropic domain growth was 
well-described by algebraic laws. 
This is, to our knowledge, the first time that 
the elliptic average shape of the domains is 
systematically investigated in a numerical experiment.
Concerning the scaling of the
the two-point correlation function,
Qiu {\it et al.}~\cite{qiu} have shown that the form (\ref{scaling1}) is 
not appropriate, but did not investigate the form (\ref{scaling2}).

Analytically, the known results in the case of a conserved
order parameter are for the $O(n)$ model, which
has been solved with shear in the large-$n$ limit~\cite{rabr}
(see also~\cite{cogola2}).
The second result stems from a scaling argument developed by 
Corberi {\it et al.}~\cite{cogola}. 
Both predict that the growth laws should be 
$R_x(t) \sim t^{4/3}$ and $R_y(t) \sim t^{1/3}$, with
a scaling form for $C(\boldsymbol{r},t)$ as in Eq.~(\ref{scaling1}).
Our results do not corroborate
these predictions.
A reason may be that these results do not take into account
the elliptic shape of the domains and we have clearly demonstrated
this could be a key point for understanding the scaling
properties of the system.
One might  also question the validity of an RG-type argument
in the case where the exponent $\alpha_\perp$ is zero, since upon
space rescaling, domains become thinner and thinner. 
Recently~\cite{brca}, an approximate solution of the case
of a non-conserved order parameter has been found, predicting 
the scaling (\ref{scaling2}) in two spatial dimensions.
However, extending this solution to the conserved
case is certainly a very hard task, since already in
the unsheared case the conservation law makes the calculations
very involved~\cite{review_alan}.

We are then faced with the problem 
of having numerical results that cannot
be understood within an existing framework.
We now give a simple physical argument leading to Eqs.~(\ref{regime}),
inspired by the original argument given by Huse~\cite{huse} to
describe the zero-shear case.
The spinodal decomposition under a  shear flow
results basically from 
two competing effects.

(i) The advection of the order parameter, which becomes
efficient for times  $t \gtrsim \gamma^{-1}$,
deforms the nearly circular domains existing at 
time $t \sim \gamma^{-1}$.
It is very easy to compute that, with the advection
only, a circular domain is deformed into an elliptic shape
with principal axes scaling at large strain as
$R_{\perp}(t) \sim (\gamma t)^{-1}$ and
$R_{\parallel}(t) \sim \gamma t$, with a tilt angle $\theta(t) \sim 
(\gamma t)^{-1}$.

(ii) The domain growth arises because of the existence of a gradient
of the chemical potential $\mu \equiv \delta F/\delta \phi$. 
This force gives rise to currents which make an interface 
of curvature $R$ move with a velocity
$\upd R/\upd t \propto 1/R^2$.
This interface motion results, in the absence of shear, to a 
coarsening of the domain structure~\cite{huse}.
Here, we modify the argument by taking into account the fact
that the structure is no longer isotropic. While
the domain growth does not affect the tilt angle
$\theta(t)$, it leads to two different 
interface velocities, namely
$\upd R_\parallel/\upd t \sim - 1/{R_\perp}^2$ and 
$\upd R_\perp/\upd t \sim + 1/{R_\parallel}^2$.

With the strong assumption that a balance can be made between
these two effects, this leads to 
the following equations for the the three parameters of the ellipse
\begin{eqnarray}
\frac{\upd R_{\perp}(t)}{\upd t} & = & -\frac{1}{\gamma t^2} +
\frac{1}{R_{\parallel}(t)^2},
\label{heuristic1}
\\ 
\frac{\upd R_{\parallel}(t)}{\upd t} & = & + \gamma -
\frac{1}{R_{\perp}(t)^2},
\label{heuristic2}
 \\
\theta (t) & = & \frac{1}{\gamma t}, 
\label{heuristic3}
\end{eqnarray}
which indeed imply Eqs.~(\ref{regime}).
Although simple and heuristic, this argument captures in fact the essence
of the coarsening process under a shear flow. 
It should correctly describe the 
domain growth when the domains are quite large. 
Otherwise, the notion of interface velocity is meaningless. 
It is quite clear however that subtleties, such as logarithmic corrections
that exits in more involved computations~\cite{brca},
will not be captured by such a naive argument.

Let us note that
when it is applied to the non-conserved case, this  argument 
leads to the growth laws $R_\perp \sim constant$
and $R_\parallel \sim t$, at leading order.
This not so far from the result $R_\perp \sim (\ln t)^{-1/4}$ and
$R_\parallel \sim t (\ln t)^{1/4}$ found in Ref.~\cite{brca}.

The second point we want to notice is that 
these equations may be used to understand the following experiment.
We have mentioned above the possibility of applying the 
shear flow a certain time after the quench, in order to have large
circular domains as initial conditions.
From Eqs.~(\ref{heuristic1})-(\ref{heuristic3}), 
we expect the following behavior.
For a relatively small strain, since the domains are quite
large, the second term in the rhs of Eqs.~(\ref{heuristic1})
is negligible and $R_\perp$ first {\it decreases}.
Thus the second term in the rhs of Eq.~(\ref{heuristic2}) becomes 
important and the growth of $R_\parallel$ is slower than
for a direct quench in the flow.
This initial behavior is qualitatively different from the previous case
and after this unusual transient, 
the growth becomes similar to the one studied previously.
This predicted behavior agrees satisfactorily with the 
corresponding numerical
experiment, as can be seen in Fig.~\ref{initial}.
Note, in particular, that the slope of the curve $R_\parallel(t)$
is minimum when $R_\perp(t)$ reaches its smallest value, as can also be 
deduced from Eq.~(\ref{heuristic2}).

\begin{figure}
\begin{center}
\psfig{file=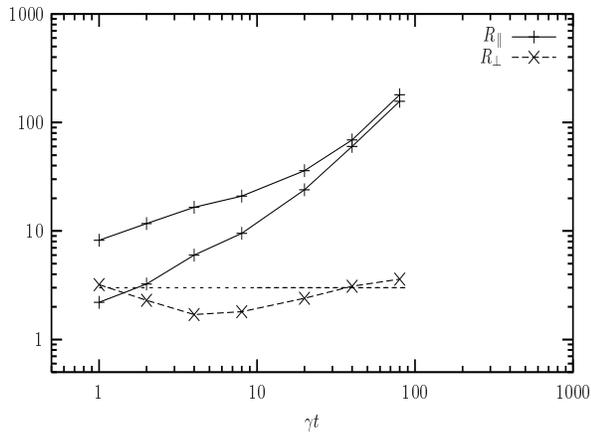,width=8.5cm,height=6.5cm} \\
\caption{Time evolution of the two length scales as a function
of the strain when
the system is first allowed to grow without shear during a time $t=400$
and then submitted to a shear flow with $\gamma=0.04$.
We have added as a visual guide a horizontal dashed line and
the data for $R_\parallel(t)$ for a direct quench in the shear flow
(lowest $R_\parallel$ curve).}
\label{initial}
\end{center}
\end{figure}

In conclusion, we hope this work has clarified
several issues concerning the spinodal decomposition in a shear flow.
It would be very interesting to perform the same
measurements in three dimensions, since the analytical approach
of Ref.~\cite{brca} shows that dynamic scaling properties might be different
in 2 and 3 dimensions.
Of course, the next (big) problem is to have a better
understanding of the effect of hydrodynamics on the phase
separation in a shear flow.

\section*{Acknowledgments}
This work was supported by the P\^ole Scientifique de Mod\'elisation
Num\'erique at \'Ecole Normale Sup\'erieure de Lyon.
It was motivated by the constant 
interest and encouragements shown by A. J. Bray and
I wish to thank him warmly for his support.
I also sincerely thank A. Cavagna for his numerous interesting 
comments and questions on this  work.
I am grateful to J.-L. Barrat and J. Kurchan for 
the continuous and fruitful interactions we had over the last two years.


\begin{references} 


\bibitem{review_hohenberg} P. C. Hohenberg and B. I. Halperin, 
Rev. Mod. Phys. {\bf 49},
435 (1977). 

\bibitem{review_gunton} J. D. Gunton, M. San Miguel and P. S. Sathni, 
in {\it Phase Transitions and Critical Phenomena}, vol. 8, edited by 
C. Domb and 
J. L. Lebowitz (Academic Press, New York, 1988). 

\bibitem{review_alan} A. J. Bray, Adv. Phys. {\bf 43}, 357 (1994).

\bibitem{review_onuki} A. Onuki, J. Phys. C: Condens. Matter {\bf 9}, 
6119 (1997), and Refs. therein.


\bibitem{beysens0} D. Beysens, M. Gbadamassi and B. Moncef-Bouanz,
Phys. Rev. A {\bf 28}, 2491 (1983).

\bibitem{beysens1} C. K. Chan, F. Perrot and D. Beysens, Phys. Rev. Lett.
{\bf 61}, 412 (1988).

\bibitem{string} T. Hashimoto, K. Matsuzaka, E. Moses and A. Onuki,
Phys. Rev. Lett. {\bf 74}, 126 (1995);
K. Matsuzaka, T. Koga and T. Hashimoto, Phys. Rev. Lett. 
{\bf 80}, 5441 (1998).

\bibitem{hobbie} E. K. Hobbie, S. Kim and C. C. Han, Phys. Rev. E {\bf 54},
R5909 (1996).

\bibitem{qiu2} F. Qiu, J. Ding and Y. Yang, Phys. Rev. E {\bf 58},
R1230 (1998).

\bibitem{lauger} J. L\"auger, C. Laubner and W. Gronski, Phys. Rev. Lett.
{\bf 75}, 3576 (1995).

\bibitem{beysens2} C. K. Chan, F. Perrot and D. Beysens, Phys. Rev. A
{\bf 43}, 1826 (1991).


\bibitem{cogola} F. Corberi, G. Gonnella and A. Lamura, Phys. Rev. Lett.
{\bf 83}, 4057 (1999); the 3D case is considered in:
F. Corberi, G. Gonnella and A. Lamura, preprint cond-mat/0009168.

\bibitem{cates} M. E. Cates, V. M. Kendon,  P. Bladon and J.-C. Desplat, 
Faraday Disc. Roy. Chem. Soc. {\bf 112}, 1 (1999). 

\bibitem{shou} Z. Shou and A. Chakrabarti, Phys. Rev. E {\bf 61},
R2200 (2000).

\bibitem{julia} A. J. Wagner and J. M. Yeomans, Phys. Rev. E {\bf 59}, 
4366 (1999).

\bibitem{zhang} Z. Zhang, H. Zhang and Y. Yang, J. Chem. Phys.
{\bf 113}, 8348 (2000).

\bibitem{yamamoto} R. Yamamoto and X. C. Zeng, Phys. Rev. E 
{\bf 59}, 3223 (1999).

\bibitem{cogola2} F. Corberi, G. Gonnella and A. Lamura, Phys. Rev. Lett.
{\bf 81}, 3852 (1998);
Phys. Rev. E {\bf 61}, 6621 (2000).

\bibitem{rothman} D. H. Rothman, Phys. Rev. Lett. {\bf 65}, 3305 (1990);
D. H. Rothman, Europhys. Lett. {\bf 14}, 337 (1991).

\bibitem{chan} C. K. Chan and L. Lin, Europhys. Lett. {\bf 11}, 13 (1990).

\bibitem{qiu} F. Qiu, H. Zhang and Y. Yang, J. Chem. Phys. 
{\bf 108}, 9529 (1998).

\bibitem{padilla} P. Padilla and S. Toxvaerd, J. Chem. Phys. 
{\bf 106}, 2342 (1997).

\bibitem{ohta} T. Ohta, H. Nozaki and M. Doi, J. Chem. Phys. 
{\bf 93}, 2664 (1990).


\bibitem{rabr} N. P. Rapapa and A. J. Bray, Phys. Rev. Lett. {\bf 83},
3856 (1999);
N. P. Rapapa, Phys. Rev. E {\bf 61}, 247 (2000).

\bibitem{brca} A. J. Bray and A. Cavagna, J. Phys. A {\bf 33},
L305 (2000); A. Cavagna, A. J. Bray and R. D. M. Travasso, 
Phys. Rev. E {\bf 62}, 4702 (2000).

\bibitem{standrews} A. J. Bray in
{\it Soft and fragile matter, nonequilibrium dynamics, metastability 
and flow}, Ed.: M. E. Cates and M. R. Evans (Institute of Physics Publishing,
Bristol, 2000).

\bibitem{rogers} T. M. Rogers, K. R. Elder and R. C. Desai,
Phys. Rev. B {\bf 37}, 9638 (1988).

\bibitem{onuki} A. Onuki, J. Phys. Soc. Jap. {\bf 66}, 1836 (1997).

\bibitem{BBK1} L. Berthier, J.-L. Barrat and J. Kurchan, Eur. Phys. J. B
{\bf 11}, 635 (1999).

\bibitem{alan_rg} A. J. Bray, Phys. Rev. B {\bf 41}, 6724 (1990).

\bibitem{huse} D. A. Huse, Phys. Rev. B {\bf 34}, 7845 (1986).

\end{references}
\end{document}